\documentclass{aa}

\usepackage{graphicx}
\usepackage{txfonts}
\usepackage{pdflscape}
\usepackage{multicol}
\usepackage[usenames]{color}
\usepackage[dvipsnames]{xcolor}

\usepackage{natbib,twoopt}
\bibpunct{(}{)}{;}{a}{}{,} 

\bibpunct{(}{)}{;}{a}{}{,} 
\makeatletter
\newcommandtwoopt{\citeads}[3][][]{\href{http://adsabs.harvard.edu/abs/#3}%
{\def\hyper@linkstart##1##2{}%
\let\hyper@linkend\@empty\citealp[#1][#2]{#3}}}
\newcommandtwoopt{\citepads}[3][][]{\href{http://adsabs.harvard.edu/abs/#3}%
{\def\hyper@linkstart##1##2{}%
\let\hyper@linkend\@empty\citep[#1][#2]{#3}}}
\newcommandtwoopt{\citetads}[3][][]{\href{http://adsabs.harvard.edu/abs/#3}%
{\def\hyper@linkstart##1##2{}%
\let\hyper@linkend\@empty\citet[#1][#2]{#3}}}
\newcommandtwoopt{\citeyearads}[3][][]%
{\href{http://adsabs.harvard.edu/abs/#3}
{\def\hyper@linkstart##1##2{}%
\let\hyper@linkend\@empty\citeyear[#1][#2]{#3}}}

\makeatother

\begin{document}

\title{\bf Asymmetrical nebula of the M\,33 variable GR\,290 (WR/LBV) \thanks{Based on observations made with the Gran Telescopio Canarias (GTC), installed at the Spanish Observatorio del Roque de los Muchachos of the Instituto de Astrofísica de Canarias, in the island of La Palma and with the Cassini 1.52-m telescope of the Bologna Observatory (Italy).
  }}
\titlerunning{Asymmetrycal nebula of GR\,290}

\author{Olga V. Maryeva \inst{1,2} \and Gloria Koenigsberger\inst{3} \and Sergey V. Karpov\inst{4,5,6} \and Tatiana A. Lozinskaya  \inst{2}  \and  Oleg V. Egorov \inst{2}
\and Corinne Rossi\inst{7,8}   \and Massimo Calabresi\inst{9} \and Roberto F. Viotti\inst{10} 
}

\institute{Astronomical Institute of the Czech Academy of Sciences, Fri\v{c}ova 298, 25165, Ond\v{r}ejov, Czech Republic, olga.maryeva@asu.cas.cz \\
\and Lomonosov Moscow State University, Sternberg Astronomical Institute, Universitetsky pr. 13, 119234, Moscow, Russia \\
\and Instituto de Ciencias F\'{\i}sicas, Universidad Nacional Aut\'onoma de M\'exico, Ave. Universidad S/N,  62210, Cuernavaca, Morelos, M\'exico, gloria@astro.unam.mx \\
\and Institute of Physics of the Czech Academy of Sciences, 18221 Prague, Czech Republic, karpov.sv@gmail.com \\
\and Special Astrophysical Observatory of the Russian Academy of Sciences, 36916 Nizhnij Arkhyz, Russia \\
\and Laboratory “Fast Variable Processes in the Universe”, Kazan Federal University, 420008 Kazan, Russia \\
\and Physics Department, Universit\`a di Roma ``La Sapienza'', Piazza le Aldo Moro 5, 00185 Roma, Italy, corinne.rossi@uniroma1.it     \\
\and INAF, Osservatorio Astronomico di Roma, Via Frascati 33, I-00077 Monte Porzio Catone, Italy \\
\and Associazione Romana Astrofili, Via Carlo Emanuele I, n$^\circ$12A, 00185 Roma, Italy, m.calabresi@mclink.it \\
\and INAF-Istituto di Astrofisica e Planetologia Spaziali di Roma (IAPS-INAF), Via del Fosso del Cavaliere 100, 00133, Roma, Italy, roberto.viotti@iaps.inaf.it \\
 }

\date{  }

\abstract{ 
{\it Context:} GR\,290 (M\,33~V0532=Romano's star)  is a luminous M\,33 object undergoing photometric variability typical for luminous blue variable (LBV) stars. It lies inside Wolf-Rayet region in the Hertzsprung-Russell diagram and
possesses a WN8 type spectrum at the light minima.
%
%
Analysis of Gran Telescopio Canarias (GTC) spectra obtained in 2016 led to the conclusion that it is surrounded by an unresolved \ion{H}{II} region formed mostly of ejected material from the central star, and disclosed the presence of a second, more extended asymmetrical emission region. \\
{\it Aims:} The aim of this paper is to further explore the structure of the nearby environment of GR\,290.  \\
{\it Methods:} Long-slit spectra of GR\,290 were obtained with three slit orientations in the visual and red spectral regions. The emission-line distribution for each slit was analyzed. \\
{\it Results:} We confirm the presence of an asymmetric \ion{H}{II} region that extends $\sim$50~pc to the south; $\sim$30~pc to the north and southeast; $\sim$20~pc to the east and northwest and  $\sim$10~pc to the west.
We also present the first spectrum to be acquired of a star belonging to the neighboring  OB\,88 association, J013501.87+304157.3, which we classify as a  B-type supergiant with a possible binary companion. 
}


\maketitle

\section{Introduction}
Evolved massive stars are generally associated with \ion{H}{II} regions which may consist of leftover gas from the star formation epoch and a combination of shells and wind ejected by the star during different stages of its evolution. In general, as soon as a massive star is born, its fast wind  interacts with the remaining gas in its vicinity, propelling it outward.  At later times, mass is lost in the form of a slow wind or shell ejections during the luminous blue variable (LBV) evolutionary stage \citep{Vink2012review}, which is then followed by the Wolf-Rayet stage \citep{Crowtherreview}  in which the more powerful wind mixes into the previously ejected  material  (see e.\,g. \citet{Lozinskaya1992}).  The final interaction occurs when the star ends its life in a supernova explosion and the ejecta violently impact the surrounding interstellar medium (ISM). 

Information on the evolutionary path of the central star may be derived from studying the chemical composition and morphology of its surrounding ISM structures \citep{Gvaramadze2010,Weis2012ASPC,MartayanLobel2016}.
Structures having a chemical composition that is enriched with nuclear processed elements point to an evolved state of the parent star. Asymmetries suggest the presence of an inhomogeneous initial ISM,  non-spherical mass ejections, or interaction with winds of a binary companion or of nearby sources \citep{Lozinskaya1992}. 

Few studies have been performed of \ion{H}{II} regions surrounding individual stars in M\,33, and fewer still of the ISM environment of an LBVs in this galaxy. Thus, the study of GR\,290 (M\,33~V0532 or Romano's star), classified as an LBV candidate \citep{romano,HumphreysDavidson}, but suspected of being a post-LBV star transitioning into the WR phase \citep{Polcaro2016}, is extremely relevant. Its long term photometric behavior includes a rise to maximum brightness ($B\sim$16.1~mag) in the 1990's and a slow decline interrupted by brief brightness enhancements \citep{Polcaro2016}. Since 2013, GR\,290 is at its historically faintest state with $V=18.7-18.8$~mag \citep{2014ATelCalabresi,GR290galaxies}.  The photometric variations are accompanied by spectral changes. Only during the visual maximum of 1992 the star presented a B-supergiant spectrum \citep{szeifert}, while in all subsequent observations its spectrum was that of a nitrogen sequence (WN) Wolf-Rayet star \citep{Sholukhova2011}. History of study of GR\,290 are summarized in the review by \citet{GR290galaxies}. 
Its wind is overabundant in helium He and  nitrogen N, and underabundant in carbon C and oxygen O, compared to the typical M\,33 chemical abundances (\citet{GR290AandA}, hereafter Paper~I). Thus, depending on the age of the system, its nearby environment should  reflect the evolving wind conditions.

\begin{figure*}
{\centering 
\resizebox*{1.9\columnwidth}{!}{\includegraphics[angle=0,viewport=85 18 1290 380,clip]{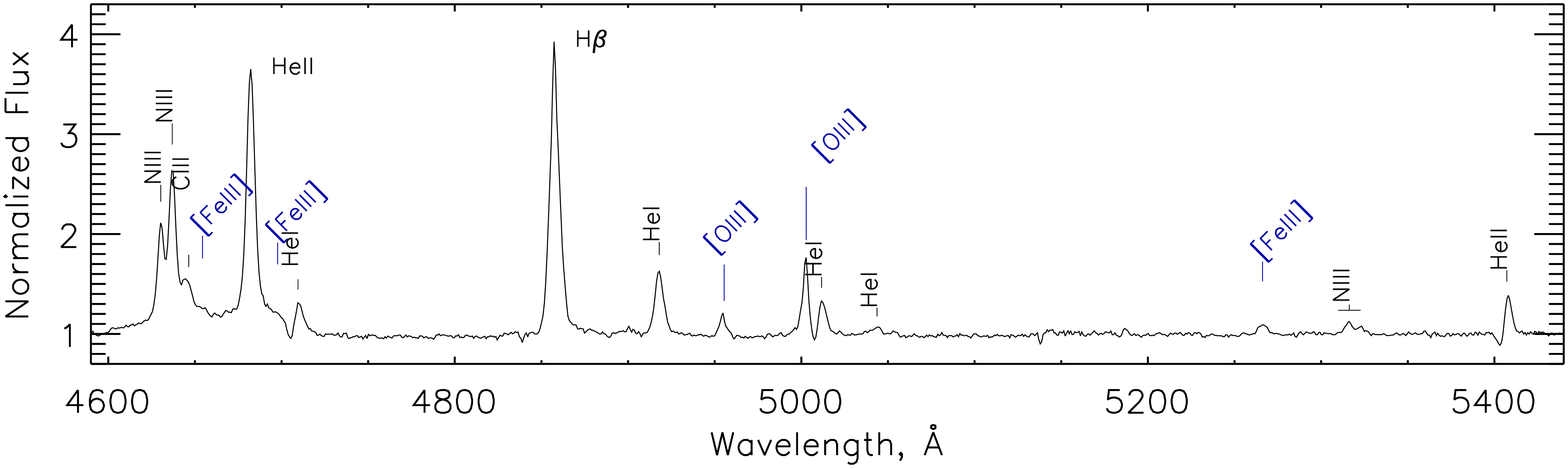}}\\ 
\resizebox*{1.9\columnwidth}{!}{\includegraphics[angle=0,viewport=85 18 1290 380,clip]{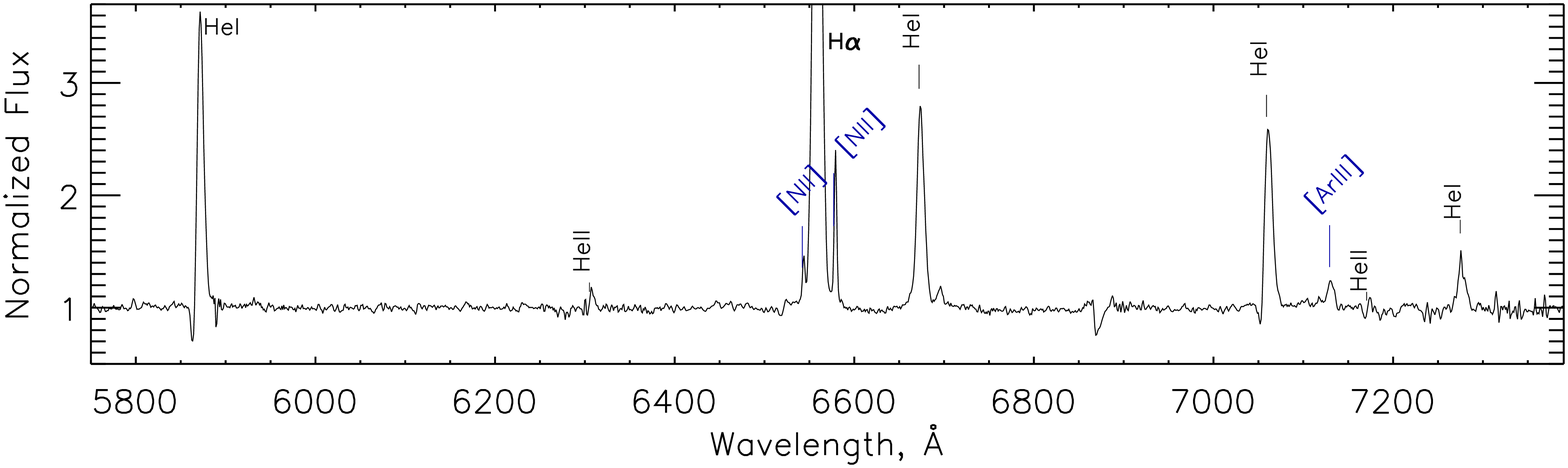}}\\
}
\caption{Normalized spectrum of GR\,290 obtained on 2018 September 8, derived by co-adding all three acquired spectra with different orientations of the slit. Principal emission lines are identified.
\label{spectrum2018}} 
\end{figure*}
\begin{figure*}
{\centering 
\resizebox*{0.95\columnwidth}{!}{\includegraphics{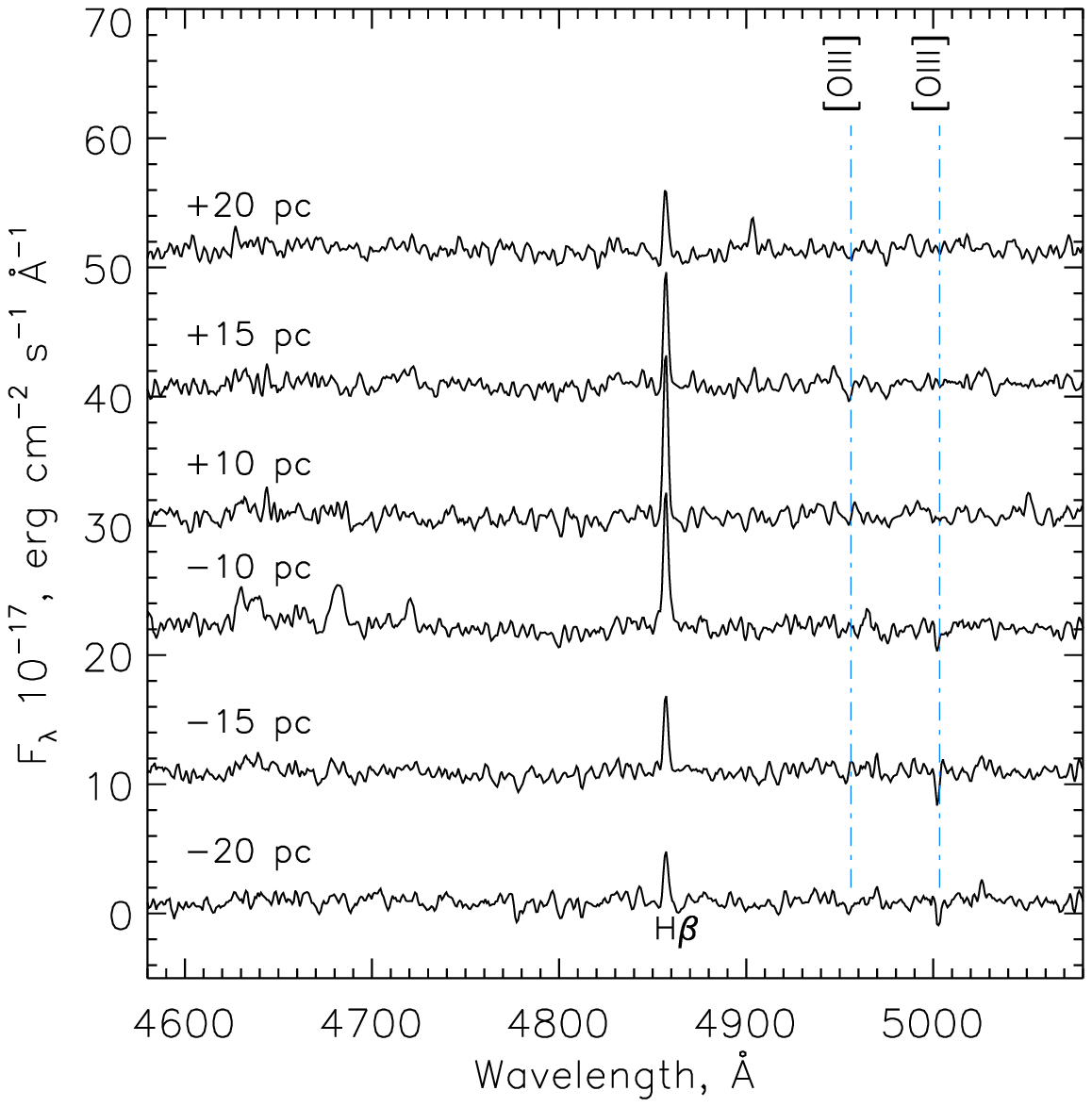}}
\resizebox*{0.95\columnwidth}{!}{\includegraphics{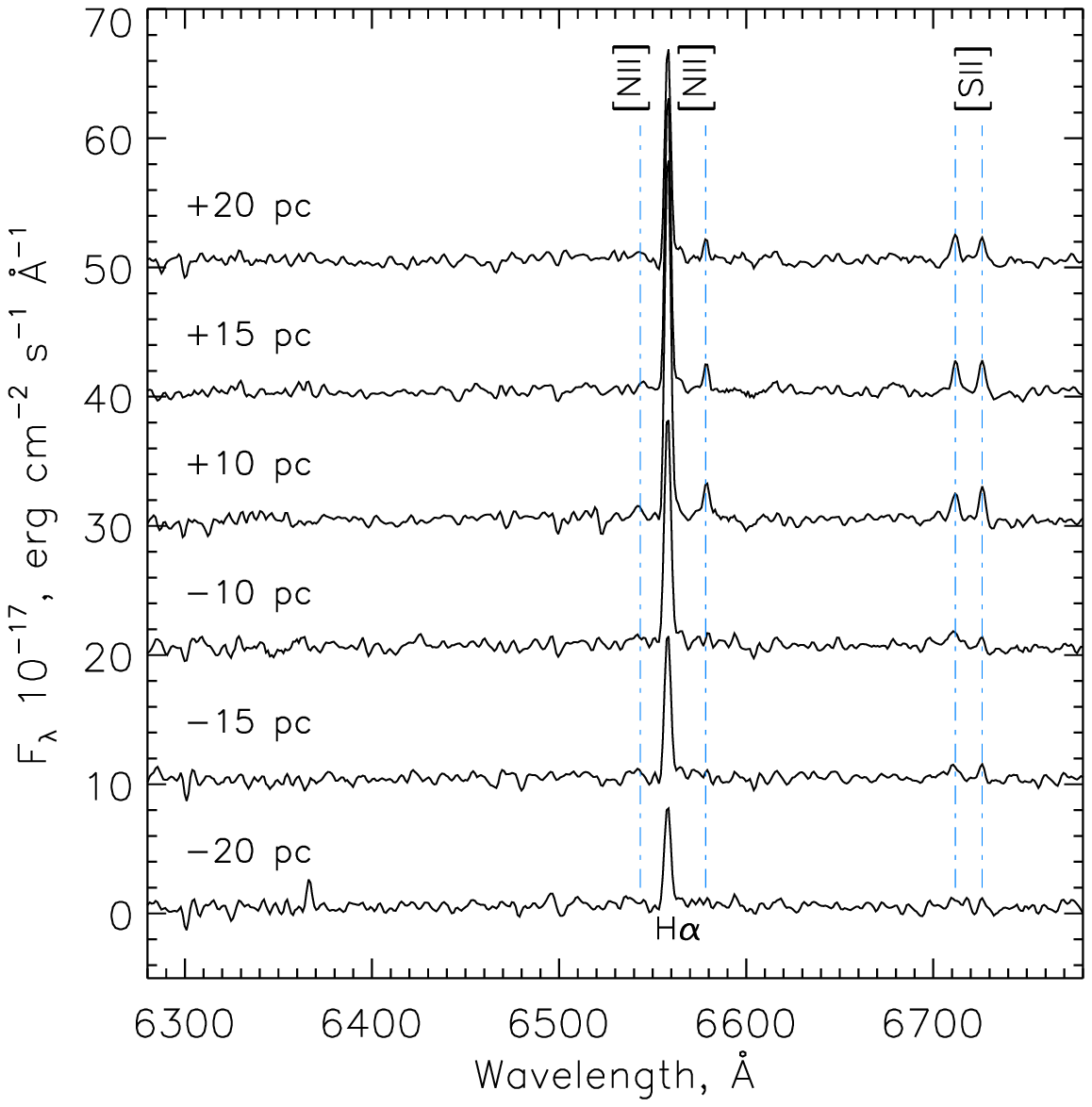}}\\
}
\caption{Emission spectra of a nebula in the regions extracted from a two-dimensional spectral image, corresponding to slit PA=135 deg (orientated Southeast-Northwest, see Table~\ref{table_obs} and Figure~\ref{fig_map}) at a different distances from the star. For clarity, the spectra are shifted vertically, original baseline of every spectrum is equal to zero. Spectra are extracted in a 5 pixels wide apertures, corresponding to the spatial resolution of 5.3 pc. The spectra are  centered on H$\beta$ (left panel) and H$\alpha$ (right panel), no other spectral region contains any detectable nebular line.
\label{nebspectra}} 
\end{figure*}

\begin{table}
\begin{center}
\begin{small}
\caption{Spectroscopic observations 2018 September 8. \label{table_obs}}
\begin{tabular}{lcc rlc}
\hline
\hline
  PA     &{\bf MJD}    & {\bf Filter ID} & {\bf $t_{exp}$} & {\bf \AA/pix} &{\bf S/N}   \\
         &             &                 &    (s)          &               &            \\
\hline 
\hline 
 0       & 58370.03991 &   R2500V        &        900      &  0.80      & 30-49      \\
         & 58370.05080 &   R2500R        &        600      &  1.035     & 18-25      \\
\\     
 135     & 58370.07376 &   R2500V        &        900      &  0.80      & 30-48      \\
         & 58370.08464 &   R2500R        &        600      &  1.035     & 25-29      \\
\\     
 90      & 58370.10314 &   R2500V        &        900      &  0.80      & 28-41      \\
         & 58370.11402 &   R2500R        &        600      &  1.035     & 21-27      \\
\hline
\hline
\end{tabular}
\end{small}
\end{center}
\end{table}

A recent analysis of long-slit spectra obtained with the Gran Telescopio Canarias (GTC) showed that GR\,290 is surrounded by a compact unresolved (0.8-4~pc) nebula with chemical abundances that, with the exception of oxygen, are similar to the current stellar wind abundances (Paper~I). Thus, this nebula originated primarily from stellar ejecta, and one might expect the presence of a larger, wind-blown bubble formed earlier during the O-star evolutionary phase. Indeed, the same study reported  a partially resolved  \ion{H}{II} region extending $\sim$11~pc to the east and $\sim$5~pc to the west.\footnote{Here and in what follows we adopt a distance of 847 kpc which leads to a scale of  1.06~pc~pixel$^{-1}$ (0.258 arcsec pixel$^{-1}$)  for GTC/OSIRIS. }  
This asymmetry is not {\it a priori} expected from a wind-blown bubble unless it is expanding into an inhomogeneus medium or asymmetric mass loss occurs.  The latter was suspected by \citet{fabrika} who reported the tentative discovery of an ISM structure centered on GR\,290 and running northeast-southwest, with a velocity field suggesting a bipolar outflow.

Interaction of GR\,290's circumstellar material with nearby stellar wind sources might also lead to asymmetries.   Although this star is located in a relatively isolated region, it is   $\sim$200~pc to the east of the OB\,89 star cluster  \citep{HumphreysSandage, Ivanov, MasseyArmandroff1995}.  This raises the possibility of an interaction between the cluster wind and that of the O-star progenitor of GR\,290.



This paper is a continuation of the work we started in Paper~I,
 where a detailed analysis of the stellar spectrum was presented, focusing on the physical properties of the star which is currently in a deep brightness minimum. In addition, the long slit spectra suggested the presence of an extended nebular region. Here we concentrate solely on the properties of its circumstellar nebula. 
Therefore, the paper presents an analysis of a spatial extent and asymmetry of nebular emission around GR\,290 based on a dedicated set of long-slit spectral observations performed with GTC/OSIRIS with three slit orientations.
In section~\ref{obs} we describe the observations and data reduction process. 
Section~\ref{results} describes the properties of the  extended nebula, which is then discussed in Section~\ref{discussion}. In Section~\ref{conclusion} we present the conclusions. 
In the appendix, we also present the spectrum of hot star J013501.87+304157.3 obtained serendipitously during our observations.

\section{Observations}\label{obs}

GR\,290 was observed with the OSIRIS spectrograph on the Gran Telescopio Canarias (GTC) on September 8, 2018 in Service Observing mode. The seeing was stable on the level of $0.8-0.9\arcsec$. The instrument was configured to use R2500V gratings for the visual (4500 - 6000 \AA), and R2500R for the red (5575 - 7685 \AA) ranges, and a long $7.4\arcmin$ slit with a width of $0.6\arcsec$. The summary of the slit position angles (PA), exposure times ($t_{exp}$), spectral resolutions and signal-to-noise (S/N) ranges is presented in Table~\ref{table_obs}.  For future reference, similar observations obtained in 2016 used a slit position of PA=94.75 deg and a seeing was $0.9-1.1\arcsec$.

The spectra were reduced using the ScoRe package\footnote {SCORE package http://www.sao.ru/hq/ssl/maryeva/score.htm} initially created for  the reduction of data acquired with the SCORPIO spectrograph\footnote{SCORPIO is Spectral Camera with Optical Reducer for Photometric and Interferometric Observations (SCORPIO) on the Russian 6-m telescope.}. It was used to perform all the standard stages of the long-slit data reduction process: bias and flat field corrections, linearization, 
flux calibration using a G191-B2B spectroscopic standard star,
 and an extraction of spectrum along the dispersion using robust Gaussian fitting, analogous to how it was done in \citet{maryeva2010}. We also removed night sky lines from two-dimensional spectral images  by interpolating them using the blank regions outside the studied objects.

\begin{figure*}
{\centering \resizebox*{1.005\columnwidth}{!}{\includegraphics{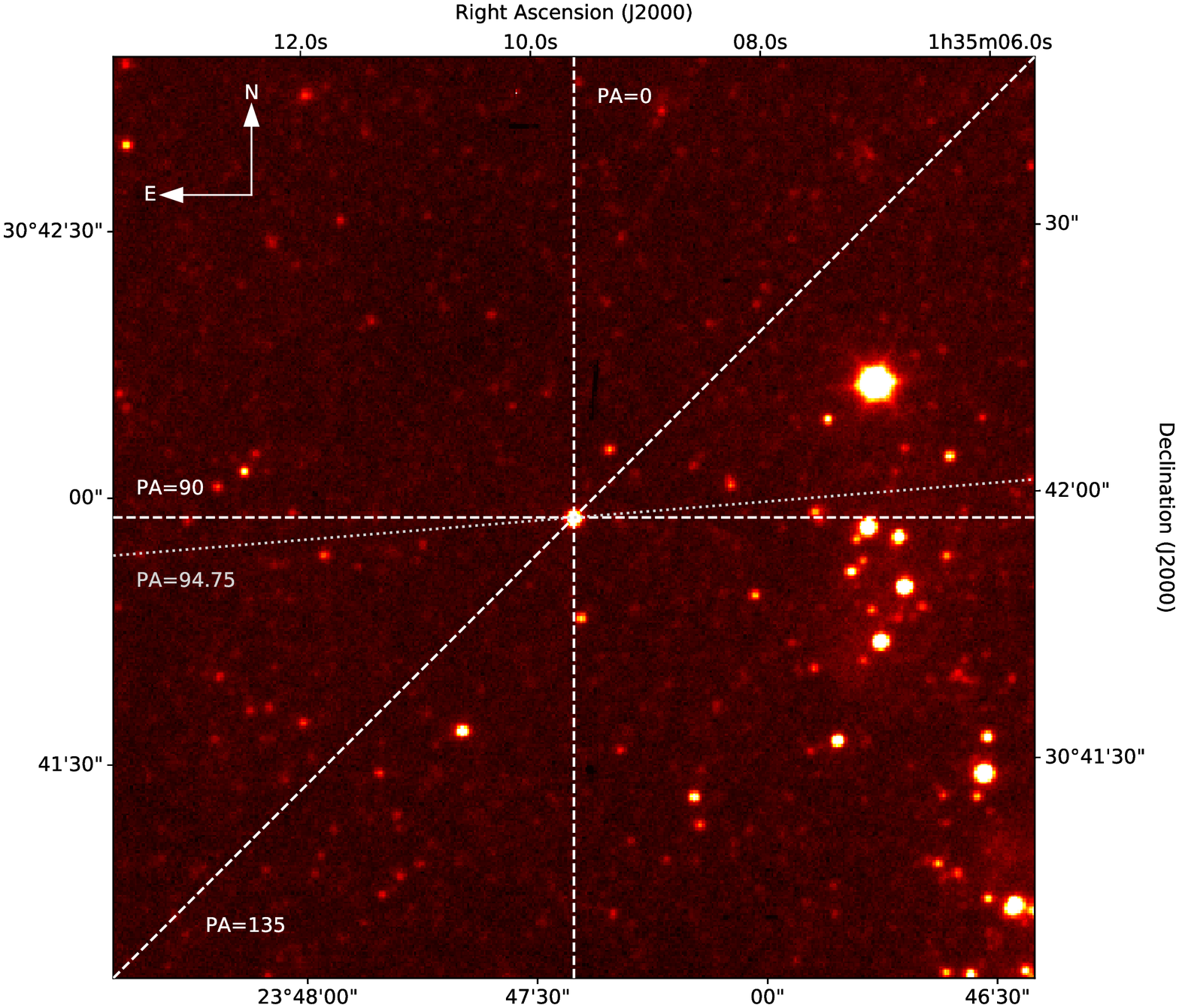}}
\resizebox*{0.995\columnwidth}{!}{\includegraphics{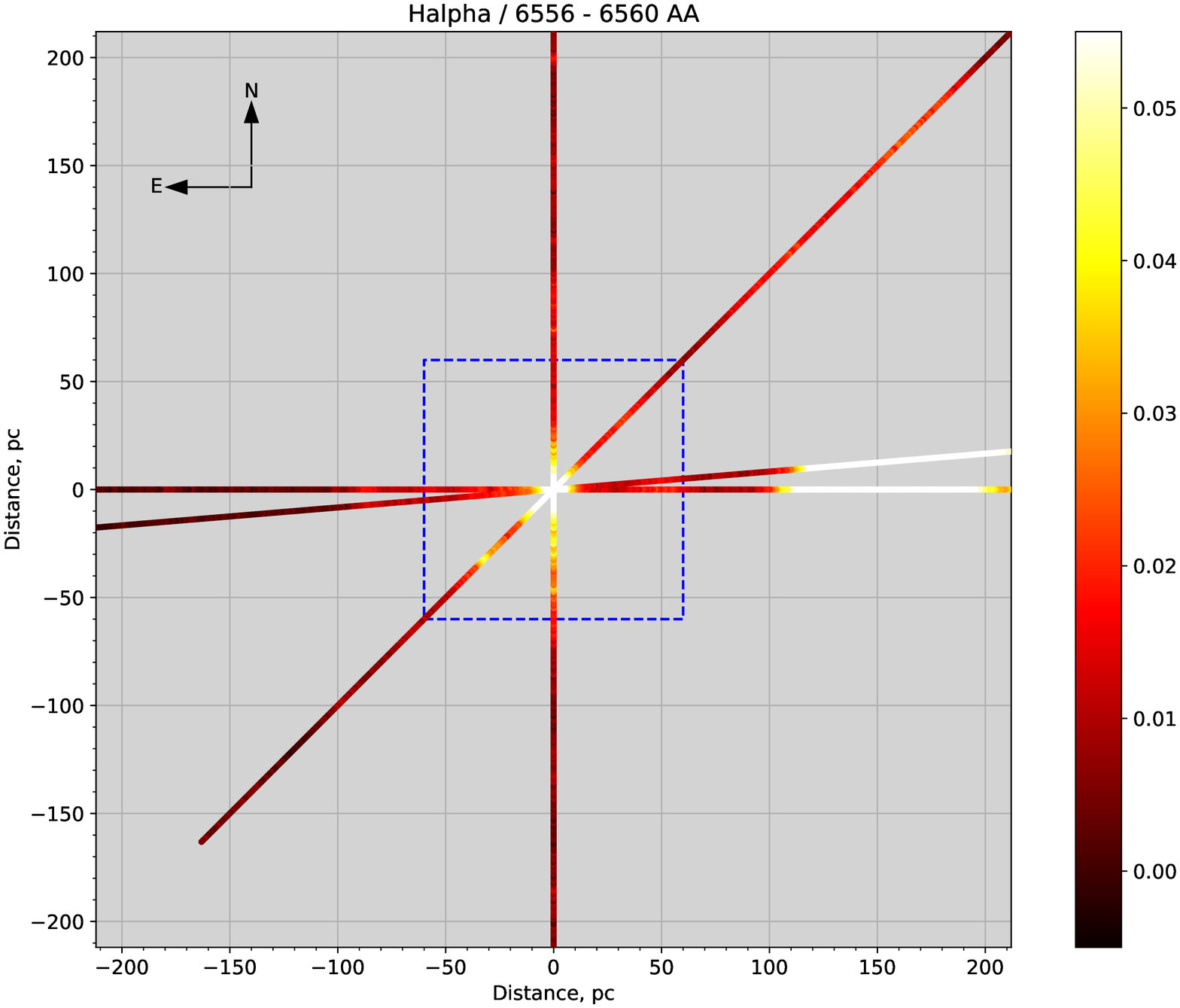}
}}\\
\caption{{\it Left:} Identification chart for a 100\arcsec\ wide region around GR\,290 in the Sloan {\it g} filter image illustrating the slit positions used during the 2018 observations (dashed lines), as well as during the 2016 (dotted line). The axes are given in RA and DEC and the position angle of each slit is listed. The stellar association OB\,89 is located to the right (West) of GR\,290.  {\it Right:}  Spatial distribution of the H$\alpha$ intensity along each of the 3 slits. The orientation and scale of the image is the same as on left panel, but here the axes are given in linear distance (pc) from GR\,290, which lies at the center of the coordinate system. Distance is computed assuming a 1.06~pc~pix$^{-1}$  scale along the slit. Strongest emission is color coded in white.  Extended emission is evident south and east of GR\,290. The dotted square defines the $\pm$60~pc sub-region that we analyze in detail.        
\label{fig_map}} 
\end{figure*}

\begin{figure*}
{\centering \resizebox*{1\columnwidth}{!}{\includegraphics{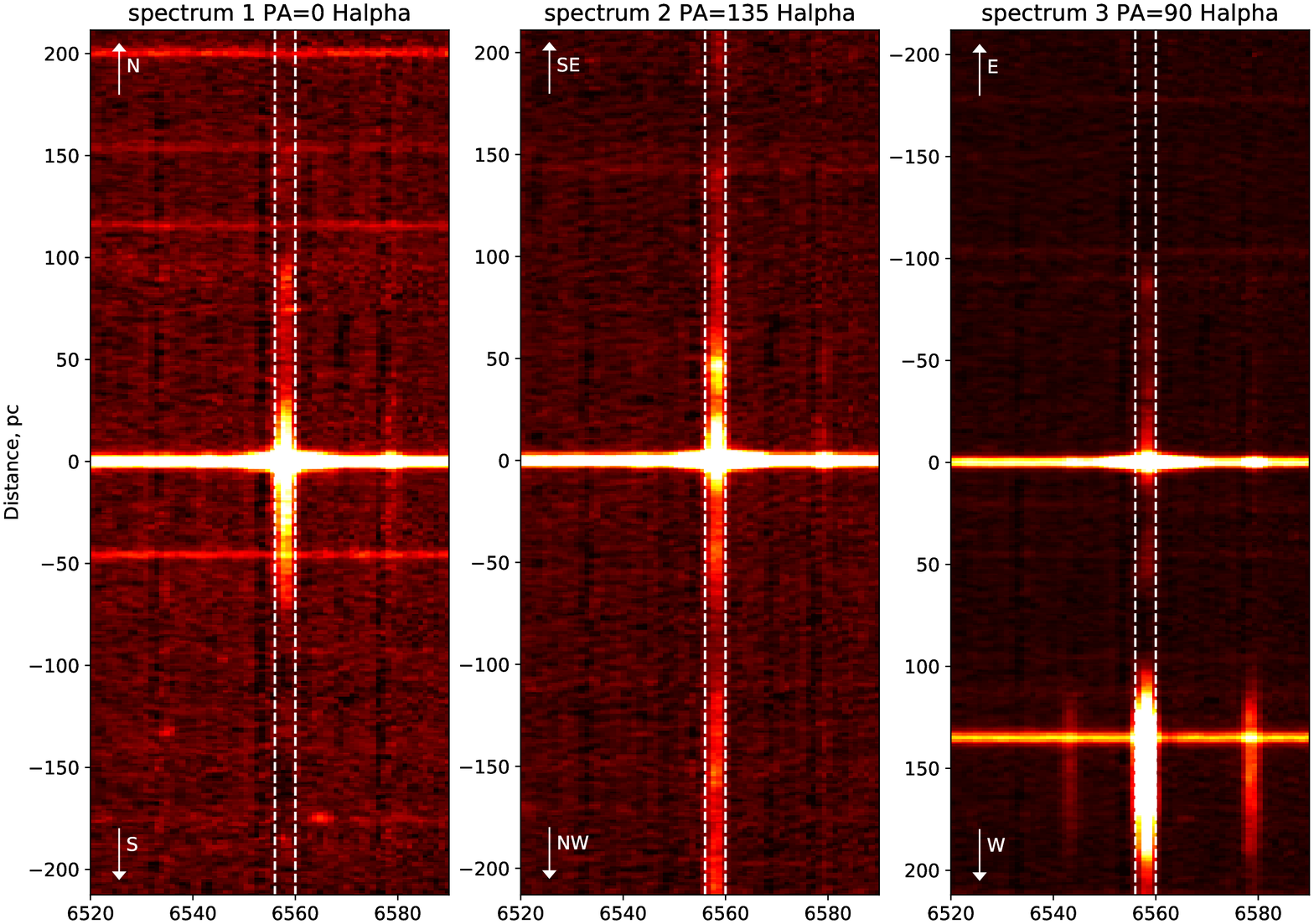}}
\resizebox*{1\columnwidth}{!}{\includegraphics{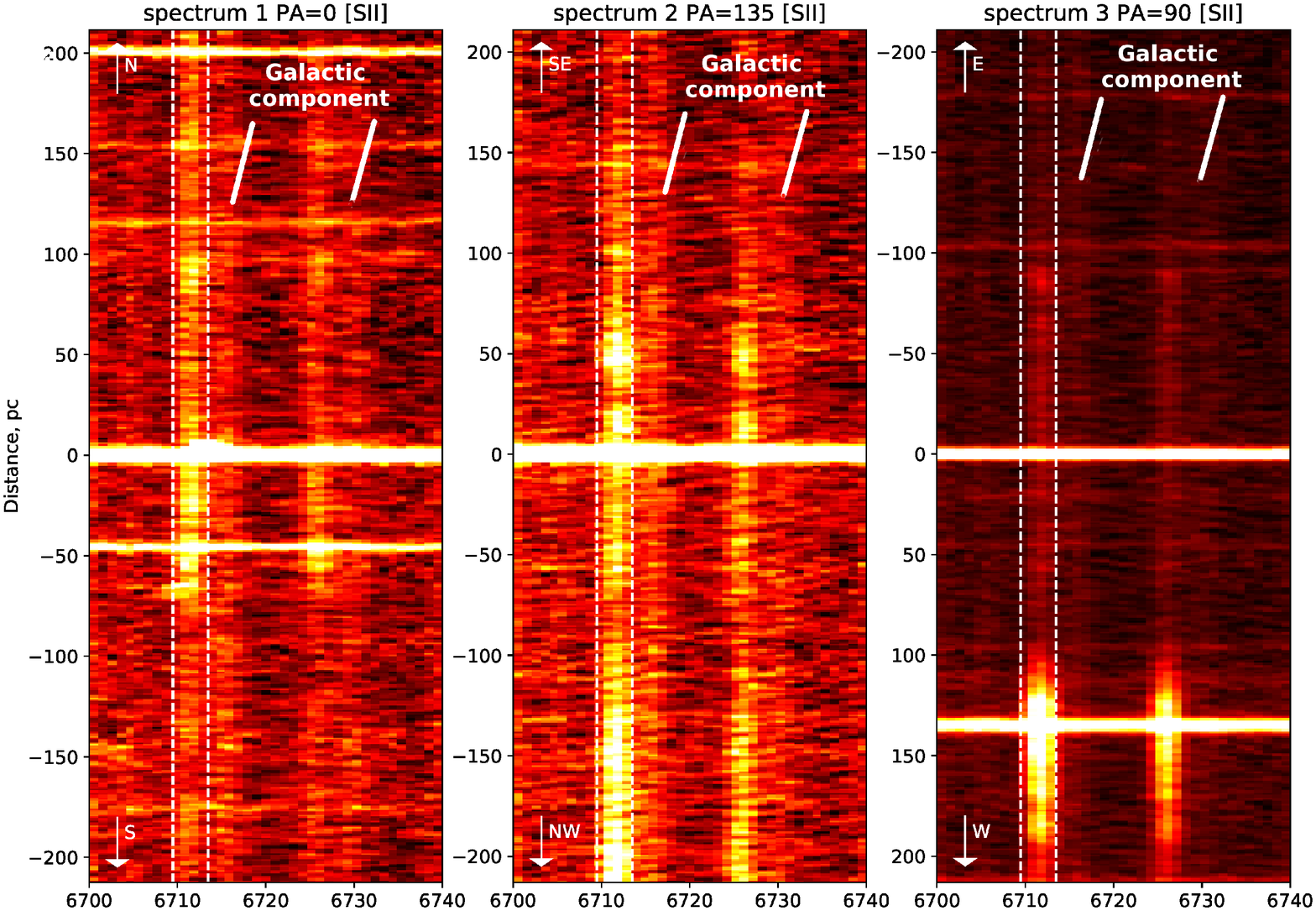}}}\\
\caption{Images showing the spatial distribution (vertical axis, given in pc) of the emission along the slit in the wavelength intervals (horizontal axis) 6520-6590 \AA\ (left panels) and 6700-6740 \AA\ (right panels), for three slit orientations.   The vertical dimension is the same as that of the dimensions in the identification chart shown in Figure~\ref{fig_map}. Arrows at the top and bottom of each image indicate the orientation.  The aperture size that was used for the extraction of H$\alpha$ and [\ion{S}{II}] along each slit is indicated with the white dash lines. In the right panels the galactic [\ion{S}{II}] emission is also visible to the right of the M\,33 emission. Dark faint traces along the vertical axis are residual marks of subtraction of night sky lines from the images, done mostly for cosmetic and presentational purposes.
\label{fig_slits_Halpha}} 
\end{figure*}

\begin{figure*}
{\centering 
\resizebox*{1.85\columnwidth}{!}{\includegraphics{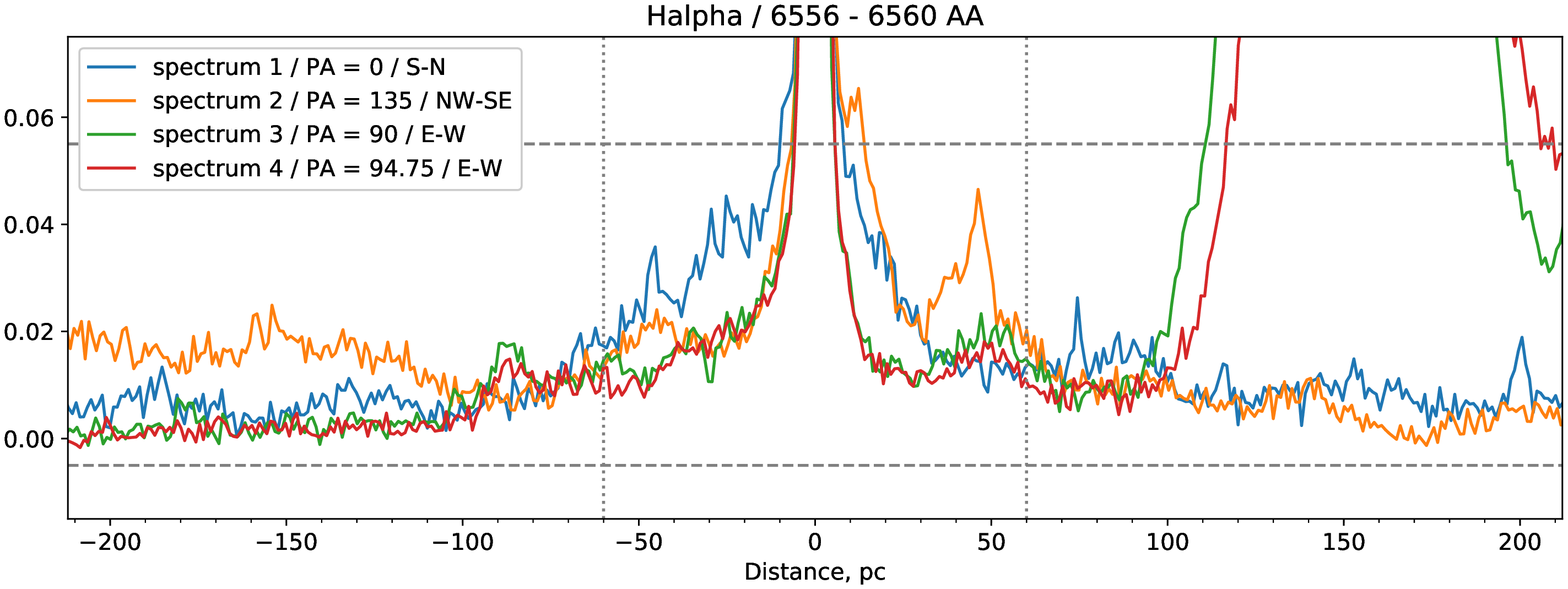}}\\ 
\resizebox*{1.85\columnwidth}{!}{\includegraphics{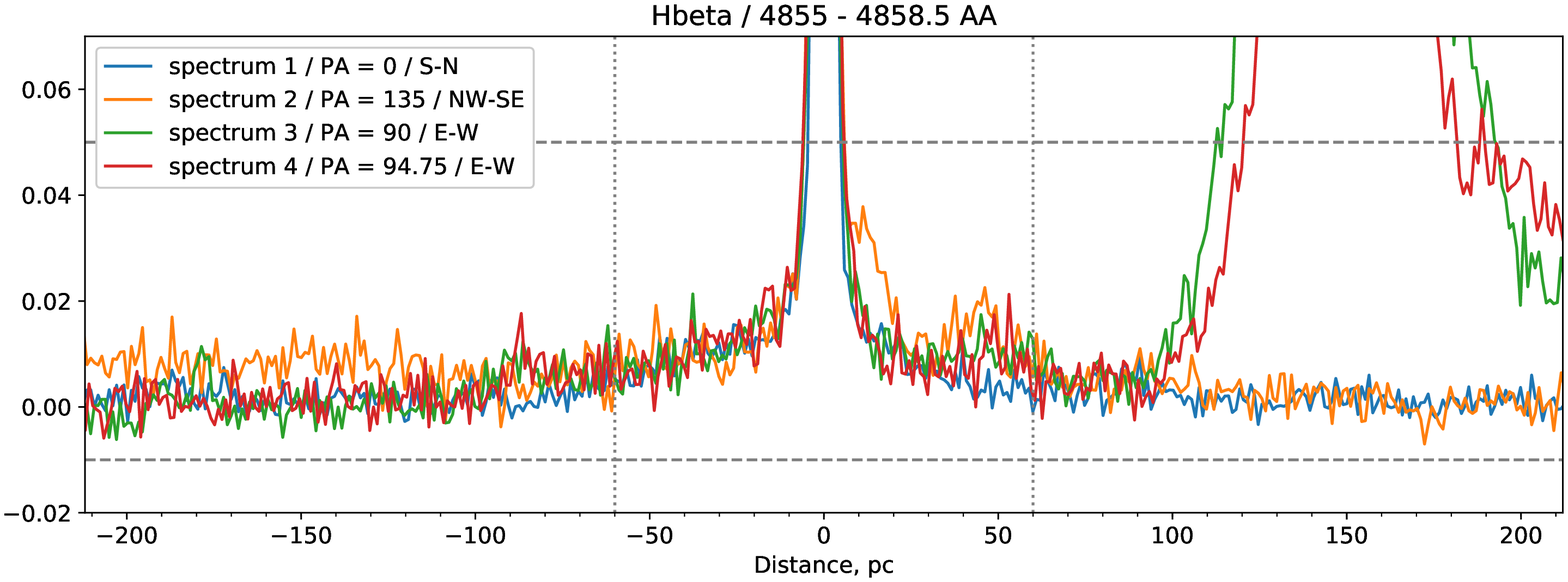}}\\
\resizebox*{1.85\columnwidth}{!}{\includegraphics{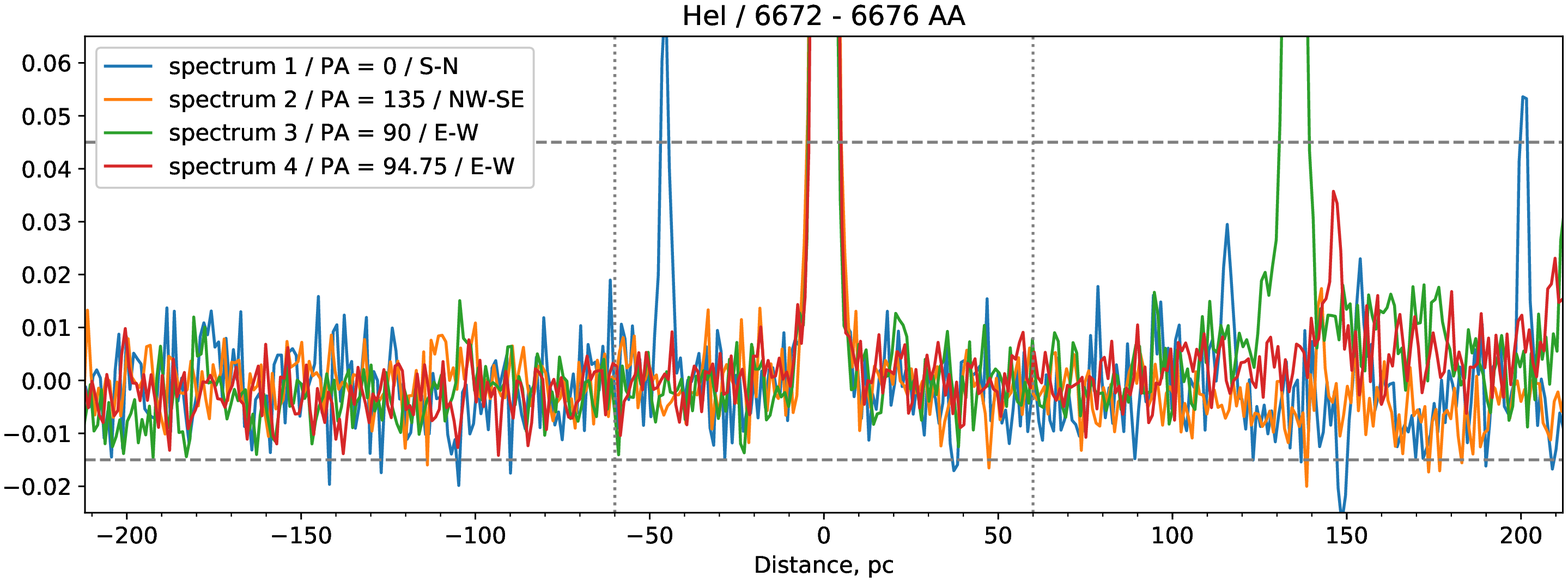}}\\
}
\caption{Tracings of the intensity of H$\alpha$, H$\beta$, 
and \ion{He}{I}  along the slit allowing a comparison of the spatial distribution of these emissions in the vicinity of GR\,290. The horizontal axis is the linear distance from GR\,290 in pc. The vertical axis gives the scaled intensity, with a scaling  such that the maximum intensity of each line is set to unity. The horizontal lines indicate the intensity interval (6\% of a peak intensity) used to define the black-to-white scale shown in Figure~\ref{fig_map} (right) and Figure~\ref{fig_region_lines}. The vertical lines enclose the $\pm$60-pc sub-region shown in Figure~\ref{fig_region_lines}.  Each panel is labeled with the wavelength interval on the observed spectrum  before correcting for the M\,33 radial velocity. The same tracings for [\ion{N}{II}] and [\ion{S}{II}] are shown below in Figure~\ref{fig_cuts_2}.
\label{fig_cuts}} 
\end{figure*}

\begin{figure*}
{\centering 
\resizebox*{1.85\columnwidth}{!}{\includegraphics{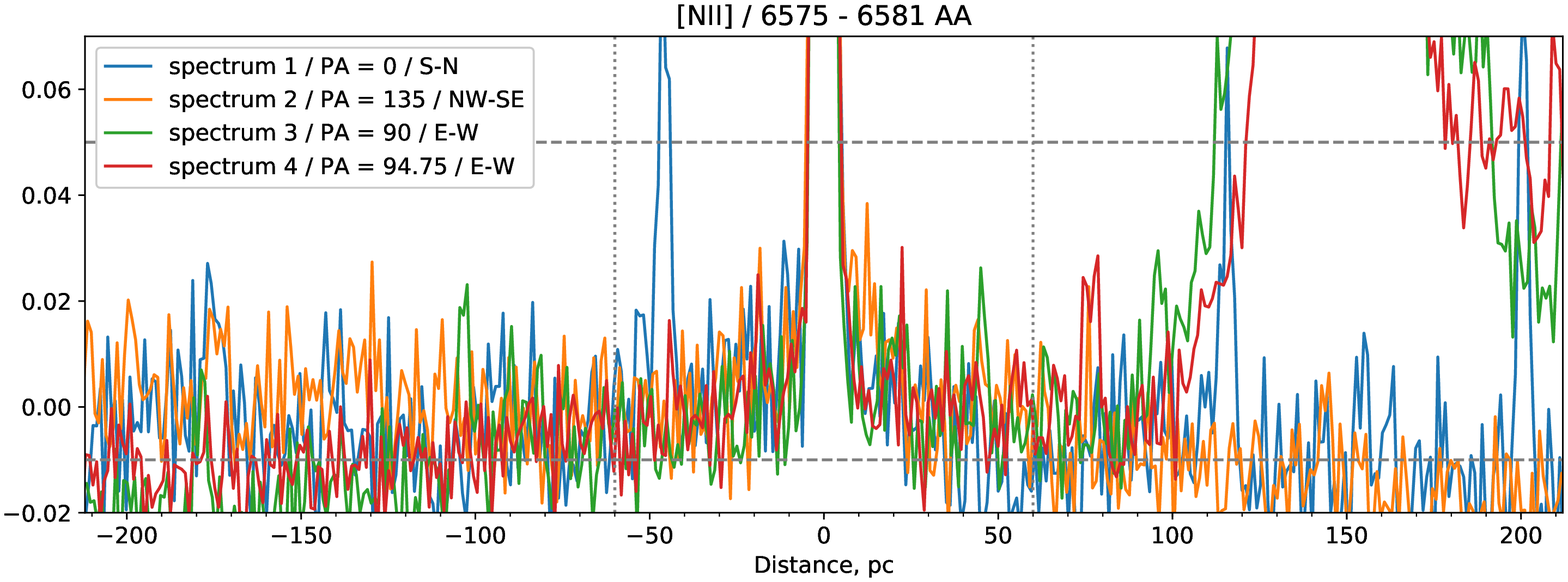}}\\
\resizebox*{1.85\columnwidth}{!}{\includegraphics{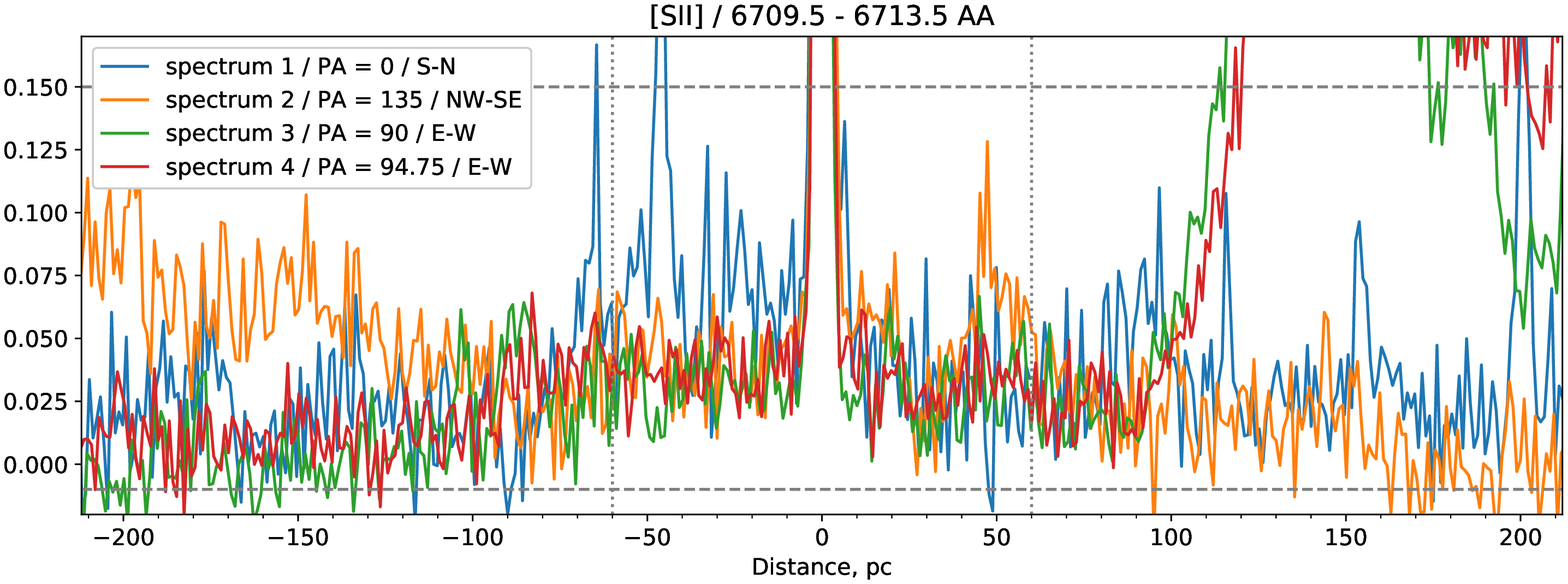}}\\
}
\caption{The same as Figure~\ref{fig_cuts} but for [\ion{N}{II}] and [\ion{S}{II}].
\label{fig_cuts_2}} 
\end{figure*}


\begin{figure*}
{\centering 
\resizebox*{0.66\columnwidth}{!}{\includegraphics{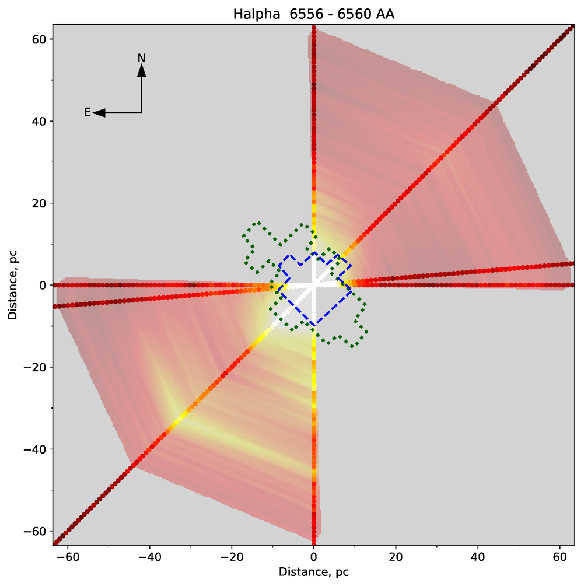}}
\resizebox*{0.66\columnwidth}{!}{\includegraphics{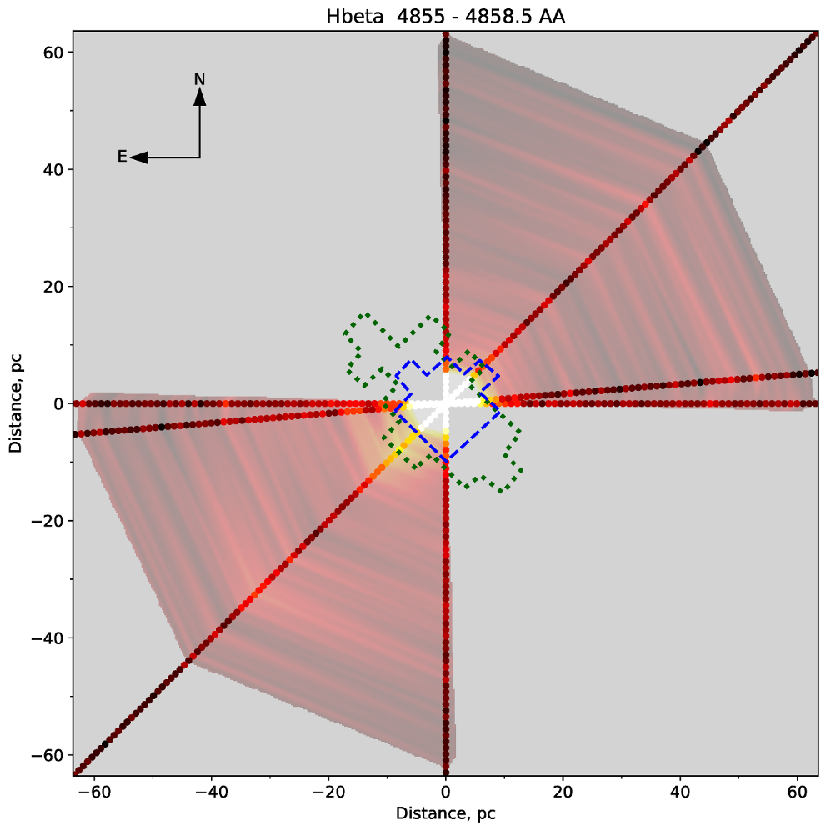}}
\resizebox*{0.66\columnwidth}{!}{\includegraphics{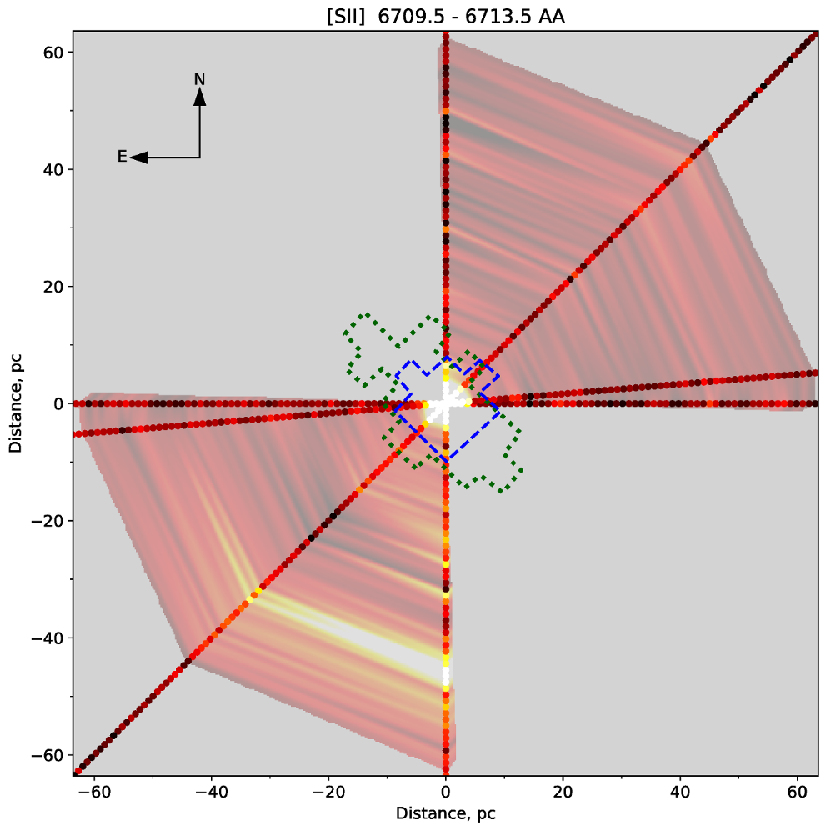}}
}\\
{\centering 
\resizebox*{0.66\columnwidth}{!}{\includegraphics{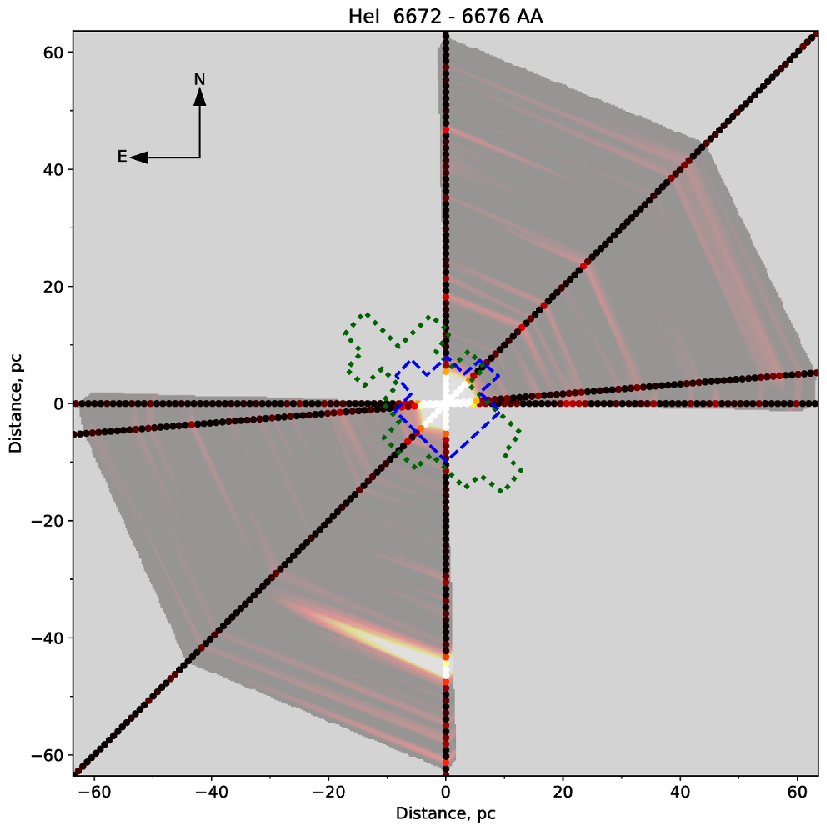}}
\resizebox*{0.66\columnwidth}{!}{\includegraphics{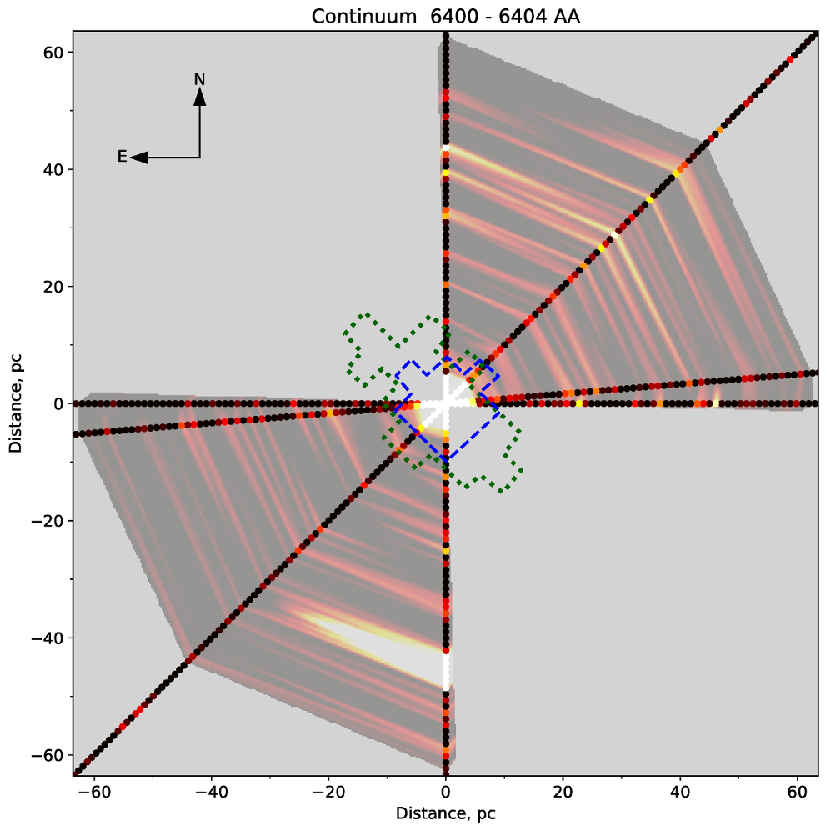}}
\resizebox*{0.66\columnwidth}{!}{\includegraphics{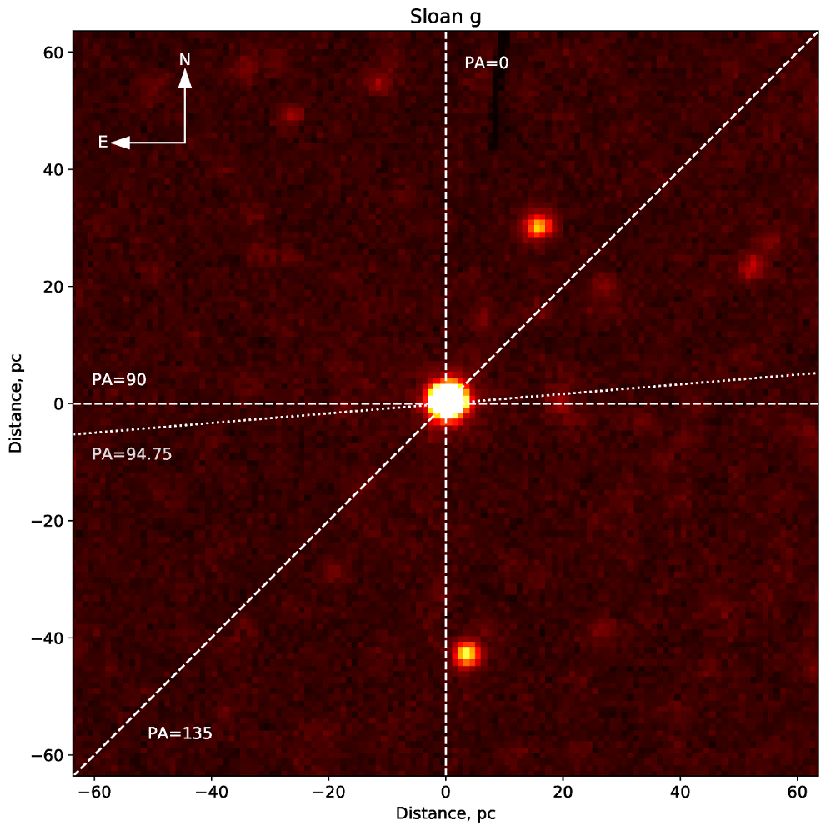}}
}\\
\caption{Spatial structure of 
the H$\alpha$ (top left), H$\beta$ (top middle), [\ion{S}{II}] (top right), \ion{He}{I} (bottom left) and a continuum at 6400\AA \ (bottom middle) emission in the $\pm$60$\times$60~pc region centered on GR\,290 according to the spatial tracings in four slit orientations. Every dot corresponds to a 1-pixel (1.06 pc) step along the slit, with black-to-white scale  defined as in Figures~\ref{fig_cuts} and \ref{fig_cuts_2} to be a 6\% of a peak line intensity (16\% for [\ion{S}{II}]). For a visual clarity, the rough linear interpolation between these points is also shown in a semi-transparent overlay inside NW-SE sectors where our coverage is dense enough. While it does not necessarily reflect the ``true'' intensity distribution between the slits and especially in NE-SW direction where we lack the data, it still perfectly depicts the significant asymmetry of the nebula along NW-SE one. Blue dashed and green dotted lines mark the extent of the nebula reported by \citet{fabrika} in narrow H$\beta$ intensity and velocity gradient, respectively.
Also, for ease of comparison, direct image of the region in Sloan {\it g} filter with the same scale and orientation is shown in bottom right panel, which is a subset of Figure~\ref{fig_map}
\label{fig_region_lines}} 
\end{figure*}

The wavelength calibration was performed using the calibration lamp spectra that were acquired on the morning following the GR\,290  observations. We find that there is a shift in the lines of the GR\,290 spectrum obtained with the PA=0 slit compared to those in the spectra obtained with the other two slits. To correct for this discrepancy, we used the telluric lines to refine the wavelength calibration.

Contemporaneous photometric observations were obtained on 2018 September 13  (MJD 58374)  using the 1.52-m Cassini Telescope run by INAF-Osservatorio Astronomico di Bologna in Loiano. The brightness of GR\,290 was: $B=18.65\pm0.04$~mag, $V=18.73\pm0.04$~mag and $R=18.63\pm0.04$~mag, confirming that the GCT observations were obtained during a deep visual minimum of the star.

\section{Results \label{results}}


The spectra of GR\,290 itself in all acquired data sets with different slit orientations are identical within the error bars. Figure~\ref{spectrum2018} shows the result of a co-addition of these three spectra.
The object spectral type is WN8h,  typical for GR\,290 at minimum brightness, and is nearly identical to the one obtained in 2016 when the star was at $V=18.77$~mag (Paper~I, \citet{GR290galaxies}). In particular, the ratio  of equivalent widths (EWs) of \ion{He}{II}{4686}/H$\beta \simeq 0.9$ remains unchanged within error bars, though the EWs of both lines were slightly smaller back in 2016. This equivalent width increase can be traced to an increase by $\sim$33$\pm$7 km~s$^{-1}$ in the line width, although the peak intensity remained constant. Except for that, overall spectral properties are practically identical to the ones described in Paper I, which corresponds to the star being in a stable deep minimum of brightness since 2013 \citep{GR290galaxies}. Therefore we will not discuss the spectral properties of the star itself further, concentrating solely on the extended nebular component.



Figure~\ref{nebspectra} shows sample spectra of the  nebula extracted at different distances from the central star for one of the slit orientations. Only H$\alpha$, H$\beta$, [\ion{N}{II}] and [\ion{S}{II}] lines are detectable there, no other line shows any spatially extended component. E.g. nebular oxygen doublet [\ion{O}{III}]~4959,~5007~\AA\ is clearly visible in the spectrum of the star itself (see Figure~\ref{spectrum2018}) but not detectable in the extended nebula, which is consistent with it being formed only in a close proximity of the star (in Paper~I we estimated the size of this inner nebula to be 0.8-4~pc). On the other hand, the spectrum of the star lacks sulphur lines [\ion{S}{II}]~6717,~6731~\AA\ apparent in the extended emission; we will specifically discuss these lines below. 
The intensity ratio of H$\alpha$ to H$\beta$ remains constant with the distance from the star and close to 3 \citep{HaHbetaratio}, which corresponds to an extended \ion{H}{II} region with negligible reddening (indistinguishable from Galaxy foreground extinction E$_{(B-V)}$=0.052 towards M33 according to the NED extinction calculator \citep{schlegel98}).



The two-dimensional spectral images obtained with GTC provide information on the diffuse emitting regions in the neighborhood of GR\,290. The slit orientations were selected to probe the ISM content along six rays extending outward from GR\,290, both coincident and perpendicular to the one used in observations of 2016 (Paper I), as well as spanning in one more direction to have better constraints on the nebular shape. 
Slit positions in the immediate 100\arcsec\ neighbourhood of the object are shown in Figure~\ref{fig_map} (left) superimposed on the direct image of the region, acquired also with GTC in the Sloan {\it g} filter. For each orientation, we extracted the emission along the slit at the positions of H$\alpha$, H$\beta$, \ion{He}{I}, [\ion{N}{II}], [\ion{O}{III}] and [\ion{S}{II}] lines using background-subtracted flux summation inside apertures 4-5 pixels (which roughly corresponds to 4-5 \AA) wide, as illustrated in Figure~\ref{fig_slits_Halpha}. The extraction apertures were defined so as to include the totality of the extended emission, with their widths and positions corresponding to the [-310:-130]~km~s$^{-1}$ range of velocities, consistent with the -179~km~s$^{-1}$ heliocentric velocity of the M\,33 galaxy. Where visible, the extended emission in these lines has a simple one-peak structure with no signs of a velocity gradient within the data precision of about 50~km~s$^{-1}$.  

The profile of H$\alpha$ emission, scaled to the maximum intensity value and color coded, is presented in  Figure~\ref{fig_map} (right), where the coordinate axes show the distance from GR\,290, computed adopting a 1.06~pc~pix$^{-1}$  scale along the slit. Its asymmetry is evident there, as PA=0~deg slit shows that emission is more extended towards the south than to the north. Similarly, the emission in the PA=135~deg slit is more extended towards the southeast and that of the PA=90~deg slit is more extended towards the east. 
Evidence of an asymmetric nebula visible in Balmer lines was previously reported in Paper~I based on observations obtained with a slit having an approximately East-West orientation (PA=94.75 deg, also shown in Figure~\ref{fig_map}). All these data show that the strongest asymmetry actually appears in the SE-NW direction (slit~PA=135 deg). 

In order to better evaluate the dimensions of the circumstellar structures, we present in Figure~\ref{fig_cuts} (top) the tracings of the scaled H$\alpha$ emission intensity along the spatial coordinate. Here we see that the resolved nebula surrounding GR\,290 extends $\sim$50~pc South; $\sim$30~pc North and Southeast; $\sim$20~pc East and Northwest; $\sim$10~pc West.  Noteworthy is the appearance of two peaks in the SE orientation, one at $\sim$15~pc and the second at $\sim$45~pc.  The latter two structures are clearly visible also in the H$\beta$ tracings shown in the second panel of Figure~\ref{fig_cuts}, confirming their reality. The more distant of these emission peaks shows up in Figure~\ref{fig_slits_Halpha}, second panel, as a separate source, although there is no stellar counterpart visible in the direct image (see Figure~\ref{fig_map}). This region is clearly extended. 

The bottom panel in Figure~\ref{fig_cuts}  shows the tracing of \ion{He}{I}~6678. This line  is  formed mainly in the stellar atmosphere. Its spatial profile, extending over $<$5~pc in each direction, is nearly identical in the three slit positions, and reflects the spatial resolution of our data. The [\ion{O}{III}] lines do not show an extended structure, while the low S/N does not allow a clear assessment of the complex structure of [\ion{N}{II}] lines (Figure~\ref{fig_cuts_2}). 

The E-W slit crosses the northern portion of the OB\,89 association, and the corresponding tracings in Figure~\ref{fig_cuts} show that the emission arising in the association  extends for approximately 100~pc, its edge being located just $\sim$100~pc from GR\,290.

The behavior of sulphur lines is significantly different from that of the hydrogen lines. [\ion{S}{II}]~6717,~6731~\AA\ lines are absent in the spectrum GR\,290 (see Figure~\ref{spectrum2018}) and we do not consider it as coming from the nebula itself. The right panel of Figure~\ref{fig_slits_Halpha}  shows that 
both of these lines have two components. Line-of-sight velocity of the blue-shifted component corresponds to the systemic velocity of the M\,33\footnote{Systematic velocities of M\,33 is -179~km~s$^{-1}$ \citep{M33velocity} and a heliocentric correction is +20~km~s$^{-1}$.}, while red-shifted component has velocity about 0~km~s$^{-1}$. The latter component has an uniform distribution of the flux along the slit. We suppose that only blue-shifted component is related to the emission from M\,33, while second component probably corresponds to foreground emission of the diffuse ionized gas (DIG) in our Galaxy which is characterised by enhanced  [\ion{S}{II}] emission (see e.g. \citet{WIM2009Haffner}). The emission from DIG of the Galaxy might contaminate the spectra of extragalactic objects in the case of their relative proximity to Galactic plane (e.g. it was clearly shown by \citet{Efremov2011} for the NGC\,6946 galaxy). 
In the further analysis we use only the blue-shifted component of [\ion{S}{II}], corresponding to the gas in M\,33. 
This component is strongly inhomogeneous and is observed in the nebula and also at different positions in its vicinity. Variations of its flux are significant, but in general, the trend with the distance from the star is absent. The [\ion{S}{II}]/H$\alpha$ ratio drops towards the star which is typical for \ion{H}{ii} regions,  but it is primarily due to the increase of H$\alpha$ intensity there.  At the distances larger than 30~pc the [\ion{S}{II}]/H$\alpha$  ratio is about 0.4-0.6 that is a typical value for DIG. 
We therefore consider these [\ion{S}{II}] lines mostly a background contribution, probably originating from the DIG in M\,33, and therefore unsuitable for additional nebular analysis.

 \section{Discussion \label{discussion}}
 
 GR\,290 is believed to be a object that has already passed through an LBV phase and is becoming a WR star \citep{polcaro10,Polcaro2016}.  It is surrounded by a compact, unresolved nebula with probable dimension $R \sim 0.8$~pc having a chemical composition similar to that of the parent star's wind,  thus suggesting that it consists of material that was ejected during the LBV phase (Paper~I). 
 
 Our new spectra  show the presence of nebular emission regions extending as far as 30~pc south, east and southeast of the star, but only 10~pc to the west. 

 To facilitate the visualization of the nebular gas spatial structure,  Figure~\ref{fig_region_lines} shows the ``line maps'' constructed using a 2D linear interpolation between the intensity points on the lines corresponding to slit positions. These maps cover the  $\pm$60~pc region centered on GR\,290 shown in bottom right panel of Figure~\ref{fig_region_lines}. The maps use the  intensity scale (black-to-white corresponding to 6\% of a peak intensity of the  corresponding line) marked with a dashed horizontal lines in Figure~\ref{fig_cuts}.  
 
 We note in Figures~\ref{fig_cuts}-\ref{fig_cuts_2} the presence of a sharp peak in both the \ion{He}{I} and [\ion{N}{II}] spatial distributions located at $\sim$50~pc south of GR\,290.  This  approximately coincides with the projected distance  of a stellar object that is visible in bottom right panel of Figure~\ref{fig_region_lines}  slightly to the west of the slit position. 
 Thus, part of the southern extension seen in H$\alpha$ and H$\beta$ could be attributed to gas that is associated with this object.  
 As the spectrum of GR\,290 lacks [\ion{S}{II}], we may conclude that its spatial structure reflects the distribution of a warm interstellar gas. Then, probably, the weak spot in  H$\alpha$ at about 20~pc towards the south from GR\,290 is also related to this warm interstellar gas.  On the other hand, there is nothing else visible within 60~pc towards the east and southeast of GR\,290, and therefore we may conclude that the spatial structure of the 50~pc emission 
 does indeed form part of an \ion{H}{II} region that is associated with GR\,290.  
 The size of the nebula around GR\,290 detected by us is close to the one around another LBV star in M\,33, Var\,2, which is most likely a result of O-star winds \citep{Burggraf2005}.  According to the estimation by \citet{Burggraf2005}, the diameter of nebula around  Var\,2 is 54~pc. 
 Thus, based on the similarity of sizes and evolutionary status of GR\,290 we may tentatively conclude  that the extended nebula around GR\,290 also resulted from a mass loss during the O-star phase.

 GR\,290 lies in an outer arm of M\,33 at a distance $\sim$17\arcmin\ from the galaxy center. In the same  arm and at a similar distance ($\sim$15.5\arcmin\ from the center, and $\sim$4\arcmin\ from GR\,290) there are two O9 stars with unusually large (diameters of $\sim$130~pc) and perfectly spherical \ion{H}{II} regions -- CPSDP\,362 and CPSDP\,364 \citep{OeyMassey1994,CPSPD1987}. Such a difference in sizes and shapes of extended \ion{H}{II} regions may be due to different ISM environments -- influences from a nearby OB\,89 association for GR\,290, and a much sparse neighbourhood of CPSDP\,362 and CPSDP\,364. Indeed, GR\,290 lies at a projected E-W distance of $\sim$200~pc to the east of the OB\,89 star association and, as can be seen in Figure~\ref{fig_cuts}, it is only $\sim$100~pc from the edge of the photoionized \ion{H}{II} region surrounding it.  Thus, the possibility exists that there is an additional ISM material beyond the photoionized  gas of the association that has inhibited the expansion of the original GR\,290 wind-blown bubble in the direction of the cluster.

\begin{figure}
{\centering \resizebox*{0.9\columnwidth}{!}{\includegraphics{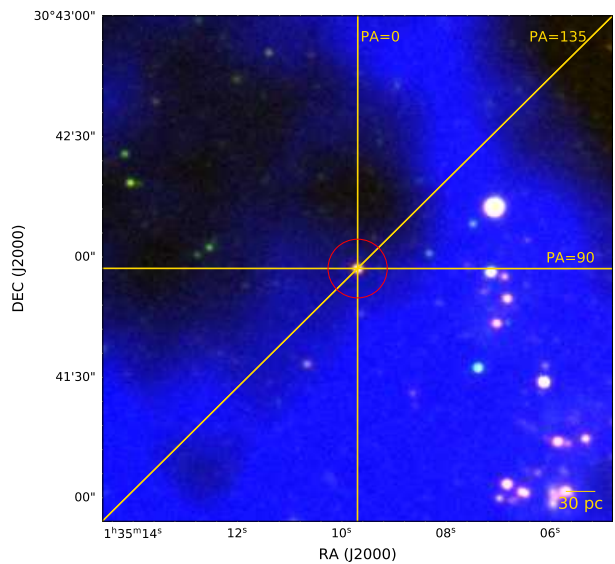}}}\\
\caption{Distribution of the neutral gas in the vicinity of GR\,290.   The color picture is a combination of three direct images, with blue channel corresponding to \ion{H}{I} 21~cm from \citet{Gratier2010H21cm}, green -- SDDS-g and red -- SDSS-r.  Position of GR\,290 is marked by a red circle with radius corresponding to 30~pc.   
\label{fig_hi21}} 
\end{figure}

 While most of the known nebulae around WR stars appear to be symmetrical, there are objects in our Galaxy that clearly display a strongly asymmetrical shape (see, e.g., \citet{Barlow1976}, \citet{DopitaLozinskaia1990}). Usually the asymmetrical shape of the nebula is explained as a result of the inhomogeneities in the surrounding ISM. That is, strong stellar wind and ejecta from WR stars blow out asymmetrical bubbles in the presence of density gradients in the surrounding ISM that are asymmetrical.
 In order to check hypothesis about the density gradients in surrounding ISM we have searched for archival data on the neutral gas distribution in this part of the M\,33 galaxy. Figure~\ref{fig_hi21} demonstrates the relative distribution of \ion{H}{I} 21~cm column density in the vicinity of the GR\,290 from \citet{Gratier2010H21cm}. It is clearly seen that while the nearby OB associations are located in the extended dense cloud of \ion{H}{I}, the GR\,290 is observed toward the edge of the local shell in the \ion{H}{I} distribution. While the influence of the wind from GR\,290 during the WR and previous stages might contribute significantly to the evolution of this shell, its size is much larger than expected for bubbles created by single WR stars. Hence, we may conclude that this \ion{H}{I} shell was created independently from the influence of the GR\,290, and because the LBV is observed toward the edge of this shell, it is very probable that its wind acts in the inhomogeneous surrounding medium.

   \citet{fabrika}  using integral field spectroscopic observations in H$\beta$  marginally resolved the circumstellar nebula of GR\,290. They found a slight excess in the narrow (with $\sim$4\AA\ FWHM) line component extending for up to 5\arcsec ($\sim$20.5~pc) from the object, and a velocity gradient across the region extending for about $\sim 30$~pc from the southwest towards northeast:
   -219~km~s$^{-1}$ in the central part (mainly the stellar velocity), -191~km~s$^{-1}$ in the SW, and -235~km~s$^{-1}$ in the NE. Thus, part of this nebula is
   approaching the observer and part is receding, as in the case of a bi-polar outflow inclined with respect to the sky plane. 

 Due to the layout of the slits used in our observations (see Figure~\ref{fig_map}) we can't confirm the extent of the nebula in the SW-NE direction, and in other directions the extent in our data is clearly detectable up to 50~pc (SE direction) from the object. On the other hand, the brightest part of H$\alpha$ and H$\beta$ extended emission components are indeed confined within $\sim$20~pc (see Figure~\ref{fig_cuts}), so the difference in nebular size may be due to better sensitivity of our observations.
Non-detection of any velocity gradient in our data is also consistent with the primary direction of the one in \citet{fabrika} data.
   

\section{Conclusions \label{conclusion}}

We present the results of new observations of the WR/LBV object GR\,290 (Romano's star) obtained with GTC in September 2018. During the observations we obtained spectra with three different position angles (PA=0, 90 and 135 degrees). Combining the results of  the new analysis of 2D spectra and the results of calculations using  CLOUDY photoionization code from Paper~I  shows that:  
 
\begin{itemize}
           
\item[--] GR~290 is surrounded by an elongated small \ion{H}{II} region which may have been formed in a previous evolutionary phase of the central star. This  region  extends $\sim$50~pc to the south; $\sim$30~pc to the north and southeast; $\sim$20~pc to the east and northwest and  $\sim$10~pc to the west.
           
 \item[--]  Nebular forbidden lines [\ion{N}{ii}] and [\ion{O}{iii}] form in an  unresolved region near the star. These observations also confirm the size estimates based on CLOUDY calculations (Paper~I), according to which the  radius of the circumstellar regions does not exceed 4~pc, with a preferred diameter being 2~pc. 
           
\end{itemize}



 Thus, new observations presented here confirm and extend the discovery of an asymmetrical nebula around GR\,290 made in Paper~I, consistent with a tentative detection by \citet{fabrika}. 
While they are not sufficient to reveal the entire morphology of this nebula,  they definitely indicate that the diffuse emission is asymmetrically distributed towards the south-southeast.  These results form a reliable basis for future observations, for instance with an integral field spectrograph, in order to estalbish the more detailed asymmetric morphology. 


%

\section{Acknowledgements}

We thank the GTC observatory staff for obtaining the spectra and Antonio Cabrera-Lavers for guidance in processing  the observations. 
Partial support for this investigation, including the travel grant for OM to visit Mexico, was provided by UNAM/DGAPA/PAPIIT grant IN103619, which is gratefully acknowledged. 
OM acknowledges support from the Czech Science Foundation GA18-05665S and Russian Foundation for Basic Research (RFBR) grant 19-02-00779. The Astronomical Institute Ond\v{r}ejov is supported by the project
RVO:67985815. 
GK acknowledges support from CONACYT grant 252499. 
SK acknowledges for support European Structural and Investment Fund and the Czech Ministry of Education, Youth and Sports (Project CoGraDS -- CZ.02.1.01/0.0/0.0/15\_003/0000437) as well as Russian Government Program of Competitive Growth of the Kazan Federal University.
TL and OE  acknowledge support from the RFBR grant 18-02-00976 and from the Program of development of M.V. Lomonosov Moscow State University (Leading Scientific School ``Physics of stars, relativistic objects and galaxies'').
OE acknowledges the Foundation of development of theoretical physics and mathematics ``Basis''.
 %
\bibliographystyle{aa}
\bibliography{GR290_GTC}

\begin{appendix}
\section{ J013501.87+304157.3 star}

The long slit used in our observations crossed a number of other stars in the vicinity of GR\,290.  For five of these, the spectral quality  was good enough (S/N$>$10) to allow the spectra to be extracted. All extracted spectra along with other spectra of stars in the vicinity of GR\,290 were published in \citet{GR290galaxies}. 
 Here we would like to highlight one of these objects -- J013501.87+304157.3 -- which attracted our attention by its radial velocity, and to show its spectrum. 

\begin{figure}
{\centering \resizebox*{0.98\columnwidth}{!}{\includegraphics{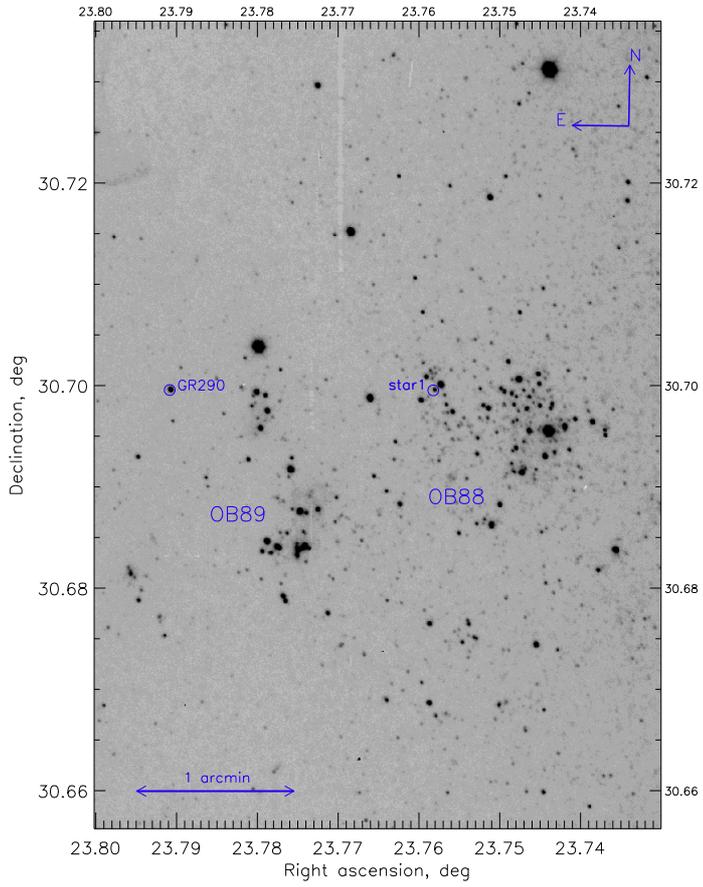}}}
\caption{Identification chart of the star J013501.87+304157.3 marked as ``star\,1''. The image was acquired with GTC in the Sloan {\it g} filter.      
\label{mapB5I}} 
\end{figure}

\begin{figure*}
{\centering \resizebox*{2\columnwidth}{!}{\includegraphics{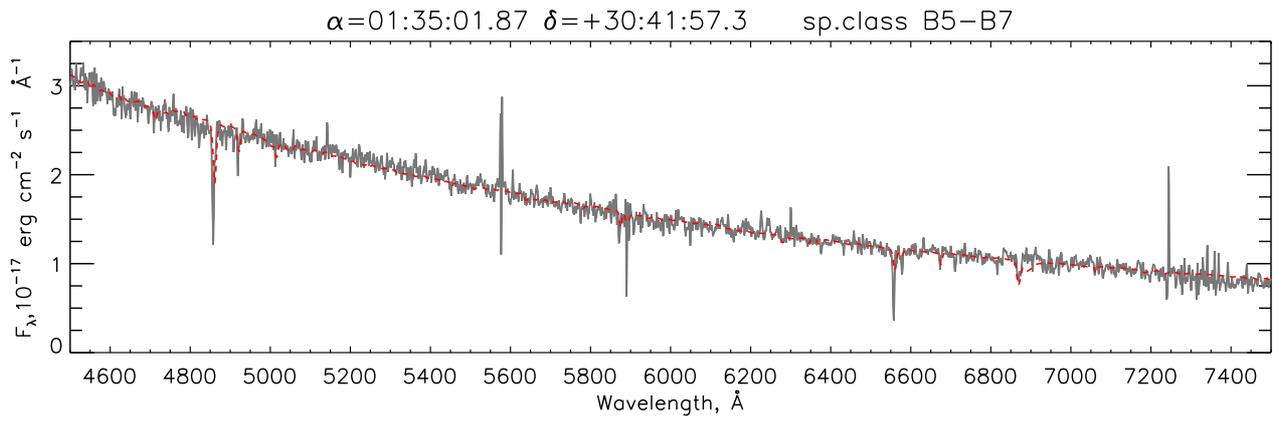}}}
\caption{Spectrum of J013501.87+304157.3. For comparison the reddened
spectra of HD\,164353 (standard star, typical example of B5\,Ib spectral type) is shown by red dashed line. \label{star1}}
\end{figure*}

This star is located in the OB\,88 association and has previously been studied photometrically by \citet{Massey2006}. Its identification chart is shown in 
Figure~\ref{mapB5I}. 
The object displays a hot early type spectrum, shown in Figure~\ref{star1}. To perform the spectral classification we used an automatic code based on the $\chi^2$ fitting with spectral standards from STELIB5 \citep{stellib} in the same way as used by \citet{CygOB2no12}. According to it, the preliminary  classification of this star is B5~I.
For comparison, Figure~\ref{star1} also shows the spectrum of a HD\,164353 standard star, which is a typical example of B5 supergiant.

H$\alpha$ and H$\beta$ absorptions at a velocity of -50~km~s $^{-1}$ relative to M\,33 rest-frame one \footnote{Velocities are corrected for the adopted M\,33 systemic velocity of -179~km~s$^{-1}$ and a heliocentric correction of +20~km~s$^{-1}$.} are prominent in the spectrum, along with a signs of a second components in both lines at a velocity of 
about $+$300~km~s$^{-1}$, suggesting the possible presence of a binary companion. 
We may also tentatively identify \ion{He}{I}~{5015}, {5875} and {7065} and \ion{C}{II}~{6578} and 6582 \AA\  at -7~km~s$^{-1}$, but these are significantly weaker.

\end{appendix}

\end{document}